\documentclass[12pt]{iopart}
\usepackage{hyperref} \usepackage{iopams}
\usepackage{epsfig}

\usepackage{graphicx}

\def\lsi{\raise0.3ex\hbox{$<$\kern-0.75em\raise-1.1ex\hbox{$\sim$}}}
\def\gsi{\raise0.3ex\hbox{$>$\kern-0.75em\raise-1.1ex\hbox{$\sim$}}}
\newcommand{\ls}{\mathop{\lsi}} \newcommand{\gs}{\mathop{\gsi}}

\begin{document}

\title{``Old'' Locked Inflation}

\author{Yang Liu\dag\ \ddag\,  Yun-Song Piao\dag\, Zong-Guo Si\ddag\ }

\address{\dag\  College of Physical Sciences, Graduate University of
Chinese Academy of Sciences, Beijing 100049, China}

\address{\ddag\  Department of Physics, Shandong University, Jinan,
  250100, China}

\ead{liuyangbyf@mail.sdu.edu.cn
\\~~~~~~~yspiao@gucas.ac.cn
\\~~~~~~~zgsi@sdu.edu.cn}

\begin{abstract}
In this paper, we revisit the idea of locked inflation,
which does not require a potential satisfying the normal
slow-roll condition, but suffers from the problems
associated with ``saddle inflation''. We propose a
scenario based on locked inflation, however, with an
alternative evolution mechanism of the ``waterfall field''
$\phi$. Instead of rolling down along the potential, the $\phi$
field will tunnel to end the inflation stage like in old
inflation, by which the saddle inflation could be avoided.
Further, we study a cascade of old locked inflation,
which can be motivated by the string landscape.
Our model is based on the consideration of making locked
inflation feasible so as to give a working model without slow roll;
It also can be seen as an effort to embed
the old inflation in string landscape.


\end{abstract}

\vskip 38mm
\section{Introduction}

Recently, substantial efforts have been invested in understanding
how to embed inflationary models in supergravity and the string
landscape,
e.g.\cite{Linde:2007fr,McAllister:2007bg,Bassett:2005xm}.
However, since the moduli fields which is naively expected to be
natural candidate of inflaton fields generally have protected
masses $m \sim H$, and the required slow roll condition $\eta \sim
m^2/H^2 \ll 1$, ($m$ is the mass of the field and $H$ the Hubble
parameter during inflation) is violated.
This $\eta$-problem is encountered in attempts to embed inflation
in the string landscape. Recently, Dvali and Kachru
\cite{Dvali:2003vv,Dvali:2003us} have proposed locked inflation.
It is slightly similar to hybrid inflation\cite{Linde:1993cn}, but
here the slow roll constraint is not required. In locked
inflation, the oscillations of a scalar field will trap the
``waterfall field" at the top of its saddle point for a period.
The potential energy at the saddle then drives inflation. This
model, if consistent, overcomes the hurdles faced by slow roll
inflation in supergravity and string theories. Without the
extremely flat potential, density perturbations cannot be created
by the same mechanism as in ordinary inflationary models. A
suggestion\cite{dgz,kofman} of alternative mechanism to generate
density perturbation was adopted.

However, Easther et al. analyzed the cosmological consequences of
locked inflation\cite{Easther:2004qs}. Their work showed that a
period of saddle inflation is possible to follow the locked
inflationary era, which will lead to some problems. To avoid
this disastrous outcome associated with saddle inflation, strong
constraints on the parameter space open to models of locked
inflation must be put. Afterwards, Copeland and Rajantie extended
the investigation for locked inflation\cite{Copeland:2005ey}. They
considered many constraints arising from density perturbations,
loop corrections, parametric resonance and defect formation, and
found that it is impossible to satisfy all of these constraints
without having a period of saddle inflation afterwards. Their
conclusion was quite strong, and it seemed ``the end of locked
inflation''.

We, in this paper, propose an inflation scenario, which inherits
the merit of locked inflation while avoids its fatal problems.
This scenario has the same dynamics of the oscillating field but
an alternative mechanism of the evolution of the ``waterfall
field''. Instead of rolling down along the potential, the
``waterfall field'' will tunnel to end the inflation stage as
in old inflation\cite{Guth:1980zm}. In our model, we need a
cross-coupling $\frac{1}{2}\lambda\phi^{2}\Phi^{2}$ term in the
potential as well. But instead of at saddle point, inflation takes
place when a scalar field keeps oscillating at the bottom of a local
minimum, which means at $\Phi=0$ the ``waterfall field''
$\phi$ has a positive rather than negative mass-squared. This feature
is more natural in certain multi-field inflationary setups within
stringy landscape, where each local minimum is relative with a
cosmological constant respectively.
We assume that this barrier is easy for rapid tunneling.
However, due to the existence of the coupling constant, the
non-zero vacuum expectation value of the oscillating $\Phi$ field
produces an additional mass term
$\frac{1}{2}\lambda\langle\Phi^{2}\rangle$ for $\phi$, which makes
the barrier that along the direction of $\phi$ enhanced. This is
helpful to trap the field in the false vacuum for a period,
thereby a large e-foldings could be provided. As time goes by, the
reduction of the $\Phi$ amplitude lead to the attenuation of the
barrier of $\phi$, which causes $\phi$ tunneling. In this
way we could overcome the shortcoming of original locked inflation.


In section II, we start with the basic points of the locked
inflation model, and then illustrate the constraints or failure of
it. In section III, first we take a look at the relationship between
barrier and the tunneling probability, as well as the idea of time
dependent decay rate. Then, we present our inflation scenario.
We will expand our scenario and give a generalization in section IV.
In the appendix we present a concrete example to illustrate our
general ideas and analyze the time dependence of the expansion rate.
Our conclusions are summarized in section V.

\vskip 9mm

\section{Locked Inflation}

Consider two scalar fields $\Phi$ and $\phi$ with a potential the same form as normal hybrid inflation~\cite{Linde:1993cn},
but notice that here at nopoint dose this potential satisfy the standard slow-roll requirements
\begin{equation}
V(\Phi,\phi) = \frac{1}{2}m_\Phi^2\Phi^2+\frac{1}{2}\lambda\Phi^2\phi^2+\frac{\alpha}{4}\left(\phi^2-M_*^2\right)^2,
\end{equation}
where $\lambda$ is a dimensionless free parameter.
At $\Phi=\phi=0$, the ``waterfall field'' $\phi$ has a negative mass-squared $-\alpha M_*^2$.

Generally, the necessity of slow roll lies in that it guarantees the model a sufficient e-foldings and
give a scale invariant spectrum which is required by the observation.
The usual slow-roll parameter $\eta_{\Phi}$, is given by
\begin{equation}
\eta_{\Phi}=M_{P}^2\frac{V''}{V}=\frac{m_{\Phi}^2}{3H^2},
\end{equation}
where $M_{P}=(8\pi G)^{-1/2}$ is the reduced Planck mass.
Slow roll inflation requires $\eta_{\Phi} \ll 1$, but here $m$ is of order $H$.
So $\Phi$ rolls toward its stationary point with certain speed, but overshoots and
performs oscillations near the origin which lock the $\phi$ field in its instability region.
At the beginning the oscillation energy might be comparable to the false vacuum energy, but
it gets redshifted and the energy density will be dominated by the latter
\begin{equation}
V=V(0,0)=\frac{\alpha}{4}M_*^4.
\end{equation}
This leads to an inflationary phase with an expansion rate
\begin{equation}
H^2\approx \frac{V}{3M_{P}^2}=\frac{m_\phi^2M_*^2}{12M_{P}^2}.
\end{equation}
The redshift causes the oscillating $\Phi$ field a decreasing amplitude
\begin{equation}
(\frac{\Phi_{A}(t)}{\Phi_{initial}})^2 \sim (\frac{a_{initial}}{a(t)})^3.
\end{equation}
That is
\begin{equation}
\Phi_A(t)=\Phi_{initial}\ e^{-\frac{3}{2}Ht}.
\end{equation}

Eventually, these oscillations are sufficiently damped to allow the field point to roll off the saddle point
in the $\phi$ direction, thereby ending locked inflation.
The critical amplitude $\Phi_{C}$ is the value of the amplitude when the effective mass of $\phi$
\begin{equation}
m_{eff}^{2}\simeq \lambda \Phi_{A}(t)^{2} - \alpha M_{*}^2
\end{equation}
is reduced to zero, which means $\Phi_{C}=\sqrt{\alpha/\lambda}M_*$.
This implies that the number of e-foldings is
\begin{equation}
N \simeq \frac{2}{3}\ln\frac{\Phi_{initial}}{\Phi_C}\approx \frac{1}{3} \ln\frac{\lambda\Phi_{initial}^2}{m_\phi^2}.
\end{equation}

By plugging in some representative numbers ($M_*\sim M_{P}, M\sim Tev, \lambda\sim 1$, and $\Phi_{initial}\sim M_{P}$),
they get $N\simeq50$ or so. This allows the bubble to grow sufficiently large to contain our present horizon volume.

\vskip 2mm
The key point of this model is that the dynamics of the first
field keep the second field trapped in a false minimum, resulting
in an evanescent period of inflation. However, firstly, the
oscillation maybe spoiled. We can tell from the equation of $\Phi$
that there is an additional constraint on $m_{\Phi}$. If it is too
small($<3H/2$), $\Phi$ will be overdamped, and no
oscillation will occur. If $m_{\Phi}$ is too
large ($>10H$), one can produce $\Phi$ particles via
parametric resonance, and the kinetic energy will be drained from
the $\Phi$ field rapidly, undermining locked
inflation\cite{Easther:2004qs}. Secondly, Easther et al. also
pointed out that there should be a period of ``saddle inflation'',
which results in a strongly scale dependent spectrum that is
inconsistent with observation even lead to massive black hole
formation in the primordial universe right after the locked
inflation, even if the usual slow-roll parameter
$\eta_{\phi}=m_{\phi}^2/3H^2$ is greater than unity.

For multiple scaler field,
$\textit{P}_{\xi} = (\frac{H}{2\pi})^2 \sum(\frac{\partial
N}{\partial \phi^{\textit{i}}})^2$. As will be
argued in \S3.2, at the saddle point, where the rate of the slope
is small, a slight change in the field value along the
``waterfallfield'' direction can cause a big variation of the
e-foldings, leading to a large $\frac{\partial N}{\partial \phi}$,
and also it is the fact that there are two down hill directions
rather than one, leading to further amplifications. Hence modes
leave the horizon at the onset of saddle inflation will cause
massive black hole formation when they reenter the horizon.
One way
out of this black hole problem is to let the saddle inflation era
last long enough to move this troublesome modes outside of the
present cosmological horizon, as well as all the signatures of the
locked inflation. However, this long period of saddle inflation
needs $\eta\ll 1$, to find a way to avoid which is exactly our
work's motivation. Further, the existence of the long saddle
inflation might cause the locked inflation unphysical from the
viewpoint of the Trans-Planckian Problem. Hence, to make the
locked inflation feasible, one must find a method to avoid
bringing about saddle inflation or black hole\cite{Dimopoulos:2003ce}.
\vskip 2mm
Further, Copeland and Rajantie\cite{Copeland:2005ey} considered
various constraints arising from density perturbations, loop
corrections, parametric resonance and defect formation. In their
results, the most salient and unnatural feature is that the value
of parameter $\eta_{\phi}$ is supposed to be very large. First,
they figured out that to avoid saddle inflation, the value of
$\eta_{\phi}$ has to be no less than $10^3$; Second, the defect
formation will impose a constraint on $\eta_{\phi}$ with
$\eta_{\phi}\gtrsim10^{6}$. In addition, loop
correction leads to $\eta_\Phi \gs \eta_\phi/2$, and the condition
with parametric resonance would stop inflation during the first
e-folding unless $\eta_\Phi \ls 70$. The whole parameter space
available for locked inflation is ruled out, especially that
constraints on $\eta_{\phi}$ is too strong and it's unnatural thus
can't be satisfied. It is also clear that if saddle inflation can
be avoided by some mechanism, leaving the requirements of
parametric resonance and loop corrections, locked inflation would
be feasible. And the demand of moduli fields in supergravity and
string theory ($\eta\geq {\cal O}(1)$) would also be met.

\vskip 9mm
\section{``Old" Locked Inflation}
\vskip 3mm
\subsection{Time-Dependent Nucleation Rate}

Coleman et al. developed the qualitative and quantitative
semiclassical theory of the decay of a false
vacuum\cite{Coleman:1977py,Callan:1977pt,Coleman:1980aw}. The
vacuum of higher energy density is a stable classical equilibrium
state. However, it is rendered unstable by quantum effects by
barrier penetration, which produces bubbles of true vacuum in the
sea of false vacuum. We are talking about the formation of the
bubble, and this process takes place on scales at which
gravitational effects are negligible\footnote{As the bubble keeps
growing after its formation, the Schwarzschild radius eventually
becomes comparable to the radius of the bubble because the energy
keeps growing, at this time the effect of gravitation couldn't be
ignored anymore.\cite{Coleman:1980aw}}. It is possible to obtain
explicit expressions in the limit of small $\epsilon$, the energy
density difference between the two vacua. The decay rate per
unit volume $\Gamma$ associated with this process is given by an
expression of the form\cite{Callan:1977pt,coleman:7475}
$\Gamma\sim e^{-B}$, and the coefficient $B$ is the total
Euclidean action for the bounce
\begin{equation}
B={27\pi^{2}S_{1}^{4}\over 2\epsilon^{3}},
\end{equation}
where $S_{1}=\int \sqrt{2V} \mathrm{d}\phi$ depends on the
height, width and shape of the barrier.

Because $\Gamma$ is the exponential function of $B$, the decay rate of the false vacuum is extremely
sensitive to the parameters of the potential. As the tunneling rate is so sensitive to $S_{1}$and $\epsilon$,
the transition from a long-living field to a rapid tunneling field requires only a small change of the potential.
This could be realized by a little reduction of the height or width of the barrier as well as sort of enhancement
of the initial and final energy difference $\epsilon$.


The lifetime of the false vacuum is about
$\tau=\frac{3}{4\pi}\frac{H^{3}}{\Gamma}$\cite{Guth:1982pn,Turner:1992tz}.
The dimensionless quantity $\frac{\Gamma}{H^{4}}$ is the volume fraction of
space occupied by bubbles nucleated over a hubble time.
Old inflation required $\frac{\Gamma}{H^{4}}\ll1$, so that enough
e-foldings could be provided, but the production of bubbles of true
vacuum is rare which means bubbles could not collide with each
other and thermalization could not be achieved. And to
successfully end the inflation a general condition,
$\frac{\Gamma}{H^{4}}\geq9/4\pi$, must be
satisfied\cite{Guth:1982pn,Turner:1992tz}. People realized that to
solve the problem one needed a time dependent nucleation rate for a
single tunneling field, i.e. $\frac{\Gamma}{H^{4}}$ started out
small so that the universe inflated and later due to some
mechanism, $\frac{\Gamma}{H^{4}}$ for the tunneling field became
large so that the phase transition suddenly took place and
completely throughout the universe.

The variation of both $\Gamma$ and $H$ is frame dependent.
In extended \cite{La:1989st} and
hyperextended\cite{La:1989pn,Steinhardt:1990zx} inflation,
the Hubble constant becomes time-dependent due to
Brans-Dicke gravity\cite{Brans:1961sx}.
Later, approaches to obtain a time-dependent decay rate
utilizing two coupled scalar fields were proposed\cite{Linde:1990gz,Adams:1990ds,Copeland:1994vg}.
Adams and Freese found that in order to bring about sufficient
e-foldings the potential of the employed rolling field
must be flat just as in slow roll condition. In this paper
we go further down this route, proposing a landscape-based
model without slow-roll.

\vskip 3mm
\subsection{``Old" Locked Inflation}


In section II, we mentioned that if saddle inflation\footnote{The reason
why a period of saddle inflation takes place after locked inflation is easily comprehended.
When the constraint on $\phi$ is removed, $\phi$ stays in its local maximum with a
vanishing kinetic energy. Since the rate of the slope is small
near the local maximum, it really takes time for $\phi$ to finish
the startup. During this period, the kinetic energy is negligible
and the potential energy stays almost the same which drives saddle
inflation. Unless the rate of the slope near $\phi$ is
unbelievably large which requires $\eta$ is larger than 2000 (if $N\simeq50$\cite{Copeland:2005ey}),
this second inflation phase could not be ignored.} could be
avoided by some mechanism, locked inflation would be feasible, and
the feature of moduli fields in supergravity and string theory
($\eta\geq {\cal O}(1)$) would also be met. We find out a
different evolution of the $\phi$ field so as to turn the saddle
inflation era away\footnote{We design a different evolution of $\phi$
realized by barrier penetration so that saddle inflation could be avoided,
because after the barrier penetration, the rate of the slope of
the potential is much larger (which is different from that at the top of the local maximum), and
$\phi$ might hold an nonzero initial kinetic energy. Thus
there is no need of a long time for starting up compared with the
above condition and neither the constraints of $\eta\geq2000$.}.
By achieving this, the failure of locked
inflation talked in section II could be circumvented.

We will introduce the same dynamics of the oscillating field but
an alternative mechanism of the evolution of the ``waterfall
field''. Instead of at saddle point, inflation takes place when a
scalar field keeps oscillating at a bottom of a local minimum,
which means at $\Phi=0$ the ``waterfall field'' $\phi$ has
a positive other than negative mass-squared. We assume that this
barrier is easy for rapid tunneling\footnote{Our designation is
more natural in string landscape but would run into the
problem whether inflation could withdraw elegantly compared with
locked inflation. To make this problem settled, we
need a prerequisite that $\frac{\Gamma}{H^{4}}\geq9/4\pi$ at $\Phi=0$.
And this prerequisite may be liable to be satisfied in the potential of
multidimensional scaler fields, because among all of the many fields the
barrier of field with least resistance should be weak.}.
However, due to the
existence of the coupling constant, the non-zero vacuum
expectation value of the oscillating $\Phi$ field produces an
additional mass term $\frac{1}{2}\lambda\langle\Phi^{2}\rangle$
for $\phi$, which makes the barrier that along direction of $\phi$
enhanced. This is helpful to trap the field in the false vacuum
for a period, thereby a large e-foldings could be provided. As
time goes by, the reduction of the $\Phi$ amplitude lead to the
attenuation of the barrier of $\phi$, which causes $\phi$
tunneling. In this sense, this scenario is slightly similar to old
inflation, thus called ``old" locked inflation.

There are several benefits in our model. First, it would be more
natural to assume local minima, if the scenario is based on string
landscape. Second, saddle Inflation is avoided, thus the merit
of locked inflation can be reserved and the parameter space available
is nice. Third, it would be more
general as it includes both rolling and tunneling, however, we
don't need the potential to satisfy the slow-roll conditions at
all.

As the value of $\frac{\Gamma}{H^{4}}$ is extremely sensitive to the parameter in
the potential, a little modification of the barrier could lead to
a transition from slow to rapid tunneling.
Due to the existence of the coupling term, the barrier would be
higher where the value of $|\Phi|$ is larger, thus tunneling would
be more rare.
As a result, there would be an area inside which
$\frac{\Gamma}{H^{4}} \sim {\cal O}(1)$ and outside
$\frac{\Gamma}{H^{4}} \ll {\cal O}(1)$ (see a visual illustration
in Fig.3 in \ref{appendix:A}).

The value
of $|\Phi|$ at this edge
is denoted as $\Phi_{ep}$\footnote{Note that $\Phi_{ep}\neq
\Phi_{C}$: $\Phi_{C}$ is defined only in the locked inflation
model (It's the field point where the effective mass of $\phi$
equals to zero, or it's the junction of barrier and concave along
the direction of $\phi$); and in locked inflation we always have
$\Phi_{ep} > \Phi_{C}$.}, meaning $\phi$ is easy to be penetrated
when $|\Phi|$ is less than this value. Tunneling happens when the
time that the field point takes to move inside the easy tunneling
region ($|\Phi|\leq\Phi_{ep}$) is longer than the vacuum lifetime
$\tau$. In the process of oscillation, potential energy of
$\Phi$ ($V_{\Phi}=\frac{1}{2}m_\Phi^2\Phi_{amplitude}^2$) is
translated into kinetic energy at the lowest point with a velocity
of $m_{\Phi}\Phi_{amplitude}$. So the time $\Phi$ spends in the
easy-tunneling region is approximately
\begin{equation}
\Delta t \simeq \frac{2\Phi_{ep}}{m_{\Phi}\Phi_{amplitude}}
\end{equation}
And in this region, the time that penetrating takes is approximately
\begin{equation}
\tau=\frac{3}{4\pi}\frac{H^{3}}{\Gamma}.
\end{equation}
So, the critical amplitude of the end of the inflationary phase $\Phi_{crit}$ is
\begin{equation}
\Phi_{crit}=\frac{2\Phi_{ep}}{m_{\Phi}\tau}=\frac{8\pi\Phi_{ep}}{3m_{\Phi}}\frac{\Gamma}{H^3}
\end{equation}

There is no definite results of the calculation of the vacuum decay rate yet.
To give some rough numerical values, we need to give the initial and critical value of $\phi$.
Nevertheless, the choice of the critical value $T_{ep}$ (around where $\frac{\Gamma}{H^{4}}$ jumps from $\ll1$ to $\geq 9/4\pi$)
has great liberty, since the parameters of the least resistance barrier is not determined.
Considering that $\Phi_{ep}$ is reasonable to be at the value where the effective mass of $\phi$ arising from the coupling
term $\lambda\Phi^{2}$ is on the same scale of mass term $m_{\phi}^{2}$,
\begin{equation}
\lambda \Phi_{ep}^{2} \sim m_{\phi}^{2}
\end{equation}
for simplicity's sake, we straightforwardly choose:
\begin{equation}
\Phi_{ep} = \frac{m_{\phi}}{\sqrt{\lambda}}
\end{equation}
Upon that,
\begin{equation}
\Phi_{crit} = \frac{8\pi}{3\lambda^{1/2}}\frac{m_{\phi}}{m_{\Phi}}\frac{\Gamma}{H^3}
\end{equation}
This means the number of e-foldings is
\begin{equation}
\label{e-foldings}
N\simeq\frac{2}{3}\ln\frac{\Phi_{initial}}{\Phi_{crit}}\simeq\frac{2}{3}\ln\frac{3\lambda^{1/2}m_{\Phi}\Phi_{initial}}{8\pi m_{\phi}H
(\frac{\Gamma}{H^{4}})} \sim \frac{1}{3}\ln\frac{\lambda m_{\Phi}^{2}\Phi_{initial}^{2}}{m_{\phi}^{2}H^{2}}.
\end{equation}
we have used the condition $\frac{\Gamma}{H^{4}} \simeq 9/4\pi$\cite{Guth:1982pn,Turner:1992tz} in order to set our model
free from graceful exit problem.
To estimate the expansion factor we plug in some representative numbers.
For example, taking $\Phi_{initial}\sim M_{P}$ (see~\ref{appendix:B}
for elaboration on this point), $H\sim TeV$,
$\lambda\sim1$ and assume $m_{\Phi}\sim m_{\phi}$.
We get $N\simeq25$ or so. Note that this will enable us to reheat the universe to $10^{11} GeV$.
\vskip 2mm
Generally, for the simplest inflation model, the inflaton serves
many purposes: driving inflation, ending inflation and generating
spectrum of density perturbations which requires $\textit{P}_{\xi}
\sim H^{2}/m_{p}^{2}\epsilon \sim 10^{-5}$, this puts a constraint
on hubble scale parameter $\sim 10^{12}$ TeV (however, in modern
cosmology the most stringent constraint is from BBN, and the lower
bound on the reheating temperature would be set by baryongenesis and
the bound could be on the Tev scale corresponding to
a $H\thicksim10^{-4}eV$.). One lesson from the string
landscape, however, is that the inflationary dynamics may be not as
simple as was expected. And in our model the tasks are easier by
different fields work together efficiently: $\phi$ drives inflation,
$\Phi$ ends inflation and some other field is responsible for the
density perturbation, and the above constraint on a high energy
scale could be removed. (An interesting work\cite{Axenides:2004kb}
takes the oscillating $\Phi$ field as dark matter while the
locked $\phi$ field drives dark energy.) Actually, recent years,
Sub-eV Hubble scale inflation
scenario\cite{Allahverdi:2006ng} is hackneyed since it could be embedded within particle
physics such as MSSM flat direction inflation\cite{Allahverdi:2006iq,Allahverdi:2006cx,Allahverdi:2006we,Enqvist:2003gh,Mazumdar:2008cr},
in which the mass and couplings are well motivated and they do
not suffer through UV related issues, also there is no
trans-Planckian problem associated as the scale of inflation is
sufficiently low and that their identities might be detected in the
laboratory such as in the LHC. And people developed ways to provide
sufficient perturbation such as curvaton\cite{Lyth:2003dt,Dimopoulos:2004yb,Rodriguez:2004yc}
and modulated reheating\cite{Dvali:2003em,Kofman:2003nx} mechanisms.

To obtain a successful model of inflation one should have
60 e-foldings. As Dvali and Kachru's argument,
\begin{equation}
\textit{R}_{\textit{today}} \sim
\frac{1}{H_{*}}e^{N}\frac{T_{\textit{R}}}{T_{\textit{today}}} \sim
10^{37}cm,
\end{equation}
which is much larger than the present Hubble size
(although the e-foldings is just less than 50, the most probable
bubble size $\sim 1/H_{*}$ is large since the scale of $H_{*}$is
really small). Actually, Dvali and Kachru get about 50 e-foldings,
by choosing $m_{\Phi}\sim m_{\phi}\sim 10^{-9}eV$, which is
protected from radiative correction by supersymmetry as Copland
et.al. pointed out. This is the reason why
we chose a very conservative value which achieve only 25 e-foldings
and enable us to reheat the universe to $10^{11}GeV$. If regardless
of this problem, we could also get much more e-foldings in one process by
choosing a much lower energy scale, but we do not
need every stage to accomplish the whole heavy task since our
model is based on stringy landscape, and naturally there should be
lots of stage of the process, and the e-foldings is the summation of each
stage's which would be sufficient. This can be seen in next
section.

Moreover, in our model the choice of the value of
$\Phi_{ep}$(i.e. $\Phi_{crit}$) is in our disposition. Because the
vacuum decay rate is extremely sensitive to the barrier parameter
and even slight extra mass term arising from coupling
($\sim\lambda\Phi^{2}\phi^{2}$) could lead to the mutation of the
decay rate if $\frac{\Gamma}{H^{4}}$ for the barrier without this
additional mass term is really close to the mutation value. In
this case $\phi_{ep}$ could be very tiny theoretically, which
might provide a sufficient e-foldings (see \ref{appendix:A} as an example).

\vskip 4mm
\vskip 4mm
We also consider an arbitrary order interaction instead of the
quartic cross-coupling
\begin{equation}
\lambda\phi^{2}\frac{\Phi^{n}}{M_{P}^{n-2}},
\end{equation}
and for the purpose of contrast we still choose
$\lambda\frac{\Phi_{ep}^{n}}{M_{P}^{n-2}} = m_{\phi}^{2}$ to give
the value of $\Phi_{ep}$. Then the form of the number of e-folding
is
\begin{equation}
N \simeq \frac{2}{3}\ln\frac{\lambda^{\frac{1}{n}} m_{\Phi} \Phi_{initial}}{(M_{P}^{n-2}m_{\phi}^{2})^{\frac{1}{n}}H}.
\end{equation}
When plotting the function, we can see that $N$ is the decreasing function of $n$ and $m_{\phi}$.
We can also write it as
\begin{equation}
N \simeq \frac{1}{3}\left[\ln\frac{\lambda
m_{\Phi}^{2}\Phi_{initial}^{2}}{m_{\phi}^{2}H^{2}} +
\frac{2-n}{n}\ln
(\frac{\lambda^{\frac{1}{2}}M_{P}}{m_{\phi}})^{2}\right],
\end{equation}
where the first term is exactly the total e-foldings in the case
of $n=2$ (see Eq.(\ref{e-foldings})), and the second term shows that the larger
$n$ we choose, the smaller e-folidings we get. Plugging in the
same representative numbers as before ($\Phi_{0}\sim M_{P}$,
$H\sim TeV$, $\lambda\sim1$), we give some of the results: for the
case $n=4$, $N\simeq13$; while, $n=6$, $N\simeq9$; and
$n=1$,$N\simeq50$.

\vskip 9mm

\section{Generalizations}

In this section, we will give possible generalizations of our
scenario. The first subsection involves a cascade with several
stages of old locked inflation. The second subsection discusses
how the original locked inflation might work, taking into account
the factor of the possibility of penetrating the barrier in $\phi$
direction. The third subsection goes beyond the oscillation
mechanism to investigate other energy form with the effect of
locking the system in an evanescent vacuum state, particular the
thermal case with a thermal mass term.

\subsection{Cascades}
\vskip 3mm

The string landscape is rather complicated, there are many
metastable vacua with a variety of discrete cosmological
constants. Thus if we consider the string landscape, it can be
expected that there could be many processes of old locked
inflation occurring during the entire evolution. Imagine the
universe originally gets trapped in a vacuum with adequate high energy
scale, which then eternally inflates and occasionally a little portion of the universe
starts the long journey to the true vacuum. During the whole
journey, field point travels along the path of largest slope and
starts oscillation when coming across a local minimum. Then due to
the inflation, the oscillation attenuates, after which the field
point tunnels the least resistance barrier and rolls down to the
next false vacuum. In this way, one can easily envision a cascade
where $H$ is getting lower as time goes by, and our universe is
now at the foot of the hill.

Note that for the moduli fields, we always have $m \sim H$.
The e-foldings provided by the \textit{i-th} stage of inflation is
\begin{equation}
N_{\textit{i}} \simeq \frac{1}{3}\ln\frac{\lambda_{\textit{i}} m_{\Phi}^{(\textit{i})2}\Phi_{initial}^{(\textit{i})2}}
{m_{\phi}^{(\textit{i})2}H_{\textit{i}}^{2}} \sim \frac{1}{3}\ln\frac{\lambda_{\textit{i}}
m_{\Phi}^{(\textit{i})2}\Phi_{initial}^{(\textit{i})2}}{H_{\textit{i}}^{4}}.
\end{equation}
The e-foldings after \textit{i} stage of inflation will be
\begin{equation}
N=\sum N_{\textit{i}} \simeq \frac{1}{3}\ln \prod \frac{\lambda_{\textit{i}} m_{\Phi}^{(\textit{i})2}\Phi_{initial}^{(\textit{i})2}}{H_{\textit{i}}^{4}}.
\end{equation}
Considering the constraint on $m_{\Phi}^{(\textit{i})}$ and $\Phi_{initial}^{(\textit{i})}$
\begin{equation}
\frac{1}{2} m_{\Phi}^{(\textit{i})2} \Phi_{initial}^{(\textit{i})2} < V_{\textit{i}} \sim H_{\textit{i}}^2 M_P^2,
\end{equation}
to get the largest e-foldings we assume that the order of
magnitude of the fields throughout the $\textit{l}$ stages stays
the same i.e. $\Phi_{initial}^{(\textit{i})} \sim M_{P}$ as well as
$m_{\Phi}^{\textit{i}}\sim H_{\textit{i}}$. Then Eq.(20) reduces
to
\begin{equation}
N \simeq \frac{1}{3}\ln \prod \frac{\lambda_{\textit{i}}
M_P^2}{H_{\textit{i}}^{2}},
\end{equation}
which is the total e-foldings during inflation driven by a cascade
of old locked inflation.

As an example, in a situation that the ``cosmology constant'' of the $\textit{l}$ stages
are well-distributed and all the coupling parameter $\lambda_{\textit{i}}$ are the same, the total e-foldings would be
\begin{equation}
N \simeq \frac{1}{3}\ln [(\frac{\lambda M_P^2}{H_{\textit{l}}^{2} \textit{l}} )^{(\textit{l})} \textit{l} !].
\end{equation}
Again, we take $H\sim TeV$, $\lambda\sim1$, we obtain: if $\textit{l}=2, N\simeq49$; and if $\textit{l}=6, N\simeq118$.

\vskip 4mm
For the original locked inflation, if one is willing to have several
stages of the processes, the constraints would become much weaker
but not sufficient, and we illustrate it as follows.

As discussed above, in
order to avoid saddle inflation, the rate of the slope near $\phi=0$
needs to be really large, more precisely: $\eta_{\phi} \gtrsim
\frac{3}{4}N_{\textit{locked}}^{2}$, where $N_{\textit{locked}}$ is
the e-foldings achieved in one locked stage. And from the equation
of motion if $m_{\Phi}$ is too large (or $\eta_{\Phi}\gtrsim 70$),
parametric resonance will undermine locked inflation during the
first e-folding. In addition with the loop correction constraint\cite{Copeland:2005ey}
$\eta_{\Phi}\gtrsim\eta_{\phi}/2$. All together
we obtain $\frac{3}{4}N_{\textit{locked}}^{2}\lesssim
\eta_{\phi}\lesssim35$, which means the e-foldings achieved in one
stage of locked inflation should be less than 6.8.

On the other hand, in our model saddle inflation
(as well as the bothersome constraint it brought
out) is avoided by tunneling, so
the parameter space is ideal ($m\thicksim H$ is satisfied and
the ``old'' locked inflation will not
be undermined by overdamping or parameter resonance as long as
$\frac{3}{2}H<m_{\Phi}<10H$.), and there is no such constraint
of $N_{\textit{locked}}$ in each stage.
And what's more, ``old'' locked
inflation takes place in the local false vacua on landscape rather
than on saddle points of supersymmetric flat direction.
\vskip 2mm
In chain inflation\cite{Freese:2004vs,Freese:2005kt}, see also
\cite{Watson:2006px,Huang:2007ek,Chialva:2008zw,Ashoorioon:2008pj,Ashoorioon:2008nh,Chialva:2008xh}, a
cascade of tunneling events is also required for inflation.
However, here the case is different, since there exists a locked
period for the tunneling field $\phi$ due to the oscillation of
$\Phi$, in principle sufficient e-foldings for observable universe
may be obtained in each step or several steps of tunneling
cascade. While in chain inflation for each step it has
$\frac{\Gamma}{H^{4}}\geq {\cal O}(1)$, thus at least $10^2$ steps
are required to accomplish the entire inflation. In this sense,
our model seems less fine tuning. Besides, as both fast rolling
and tunneling are included in our model, it might be more
universal in a given landscape. However, if the fall of the potential
along the rolling direction is not adequate so that the oscillation
induced would be weakened, which will render the tunneling rapidly
happen in succession\footnote{This is the case that the initial
amplitude $\Phi_{initial} < \Phi_{crit}$. Instead of a transition
from $\frac{\Gamma}{H^{4}}\ll1$ to $\frac{\Gamma}{H^{4}}\sim1$,
the dimensionless parameter $\frac{\Gamma}{H^{4}}$ still remains
large, and the effect of the oscillation of $\Phi$ field is not
important.}. In this case, this tunneling series will be the same as
that of the chain inflation. Thus in realistic situation, a
combined cascade of old locked inflation and rapidly tunneling
like in chain inflation maybe more possible. However, in this
case, the e-foldings number will mainly be contributed by stages
corresponding to old locked inflation.

Our model is based on the consideration of making locked
inflation feasible so as to give a working model without slow roll;
It can also be seen as an effort to embed
the old inflation model in string landscape.

There are also other multiple stages inflation scenario, for
example, folded inflation\cite{Easther:2004ir}, multiple inflation
in string landscape\cite{Burgess:2005sb,Ashoorioon:2006wc,Ashoorioon:2008qr}, and also many inflation
scenario with multiple scalar fields, for example,
\cite{Mazumdar:2001mm,Piao:2002vf,
Majumdar:2003kd,Dimopoulos:2005ac,Easther:2005zr,Ward:2007gs,Cai:2008if,Battefeld:2008py,Battefeld:2008qg},
and inflation without slow roll\cite{Damour:1997cb}.
However, these models don't include a first order phase transition
which thus is obviously different from ours.



\vskip 3mm

\subsection{Revival of Locked Inflation}
In locked inflation model, $\Phi_{C}$ is the value of $\Phi$ when
the critical mass is equal to the minus mass. Here the effective
mass is equal to zero, i.e. when $\Phi=\Phi_{C}$ one comes to the
junction of barrier and concave. That is to say, in the outside
area where $\Phi>\Phi_{C}$ $\phi$ is restricted to the stable
point where $\phi=0$. However, in the area where $\Phi<\Phi_{C}$,
$\phi$ becomes an unstable maximum. The time $\Phi$ spends in this
stable region is that $\Delta t \simeq
2\Phi_{C}/m_{\Phi}\Phi_{0}(t)$. If $\Delta t$ is shorter than the
instability time scale $\frac{1}{m_{\phi}}$, $\phi$ is locked and
oscillation lasts, otherwise $\phi$ starts to roll down the slope.

However, in old locked inflation, we just make a little
modification to the original locked inflation. We add bulges to
the water field's potential in the original locked inflation so
that we turn the saddle point into a local minimum. And still we
require the barriers that we add to the original potential are
very weak so that they could be easily penetrated. However, when
we taking into account the possibility of tunneling, even in the
original locked inflation model there exists an interesting
possibility that the adscititious mass from the coupling
$\lambda\Phi^{2}$, could not only cancel the original negative
mass of $\phi$ but also generate a barrier in the region where
$\Phi>\Phi_{C}$, thus the tunneling of $\phi$ might have already
happened before it starts to roll, if the time $\Phi$ spends in
the region $\Phi_{C} \leq \Phi \leq \Phi_{ep}$ is longer than the
life time of the trapped $\phi$ field. This might revive the idea
of locked inflation, which, however, might not in its original
sense. We will briefly discuss this possibility here.

In the limit of large $\Phi_{amplitude}$: $\Phi_{amplitude}\gg\Phi_{ep}>\Phi_{C}$, the velocity approximately stays
the same in the region $\Phi\leq\Phi_{ep}$.
Therefore, the time $\Phi$ spends in the region $\Phi_{C} \leq \Phi \leq \Phi_{ep}$ is:
\begin{equation}
\Delta t' \simeq \frac{\Phi_{ep}-\Phi_{C}}{m_{\Phi}\Phi_{amplitude}} = \frac{\theta }{m_{\Phi}}
\end{equation}
where
\begin{equation}
\theta \equiv \frac{\Phi_{ep}-\Phi_{C}}{\Phi_{amplitude}} ,
\end{equation}
and $\theta$ lies between $0$ and $1$. If the condition
\begin{equation}
\theta \geqslant m_{\Phi}\tau = \frac{3m_{\Phi}H^3}{4\pi \Gamma}
\end{equation}
is satisfied ($m_{\Phi}\leq\frac{4\pi}{3} \frac{\Gamma}{H^4} H\simeq4H$), locked inflation is possibly feasible.
Otherwise, the troublesome saddle inflation era will follow.
Besides the constraint of $m_{\Phi}$, the outcome still depends on $\theta$ (or the value of $\Phi_{ep}$ and $\Phi_{C}$,
which are sensitive to the parameter of the potential).
And we also notice that the tunneling event occurs easier when $\Phi_{amplitude}$ attenuates to a smaller value (leads to a larger $\theta$),
which is consistent with our expectation.

\vskip 3mm
\subsection{Thermal case}

In this subsection, we discuss the thermal case,
which could also locks the system in an evanescent false vacuum
state. For instance, if the $\Phi$ field could quickly decay
(e.g., through parameter resonance effects) into quanta that
thermalize at a temperature $T_{initial}$, and the $\Phi$ quanta
come into thermal equilibrium at temperature $T_{initial}$, they
will create a thermal mass, i.e. $\sim\frac{1}{2}\lambda T^{2}
\phi^{2}$ for the $\phi$ field. 

The model in the thermal case is similar to that with the
oscillation case. Because of the entropy conservation in the
adiabatic expansion, the temperature redshifts as $Ta\propto
const.$ ($\frac{T(t)}{T_{initial}}=\frac{a_{initial}}{a(t)}$), i.e.
$T(t)=T_{initial}\rme^{-Ht}$. Thus,
\begin{equation}
N\simeq \ln(\frac{T_{initial}}{T_{crit}}).
\end{equation}
is the e-foldings number.
As in the oscillation case, to make the graceful exit problem
settled we need a prerequisite that it is
quite easy to penetrate to the outside of the local minima
when field is trapped in them without the effect of the thermal mass.

With the effect of the additional thermal mass, the barriers that trap the
field is enhanced, and $\frac{\Gamma}{H^{4}}$ decreases from lager
than $9/4\pi$ to $\ll 1$ all of a sudden. Unless the temperature
$T$ redshifts to a certain little critical value
$T_{ep}$\footnote{Note that different with the oscillation case,
here we have $T_{crit}=T_{ep}$} ,the local false vacuum is
unlikely to decay. We need the initial and critical value of $T$
before we could give some rough numerical results. And again the
choice of the critical value $T_{ep}$ (around where
$\frac{\Gamma}{H^{4}}$ jumps from $\ll1$ to $\geq 9/4\pi$) has
great liberty. We choose when the thermal mass $\frac{1}{2}\lambda
T^{2}$ is comparatives to the mass of $\phi$ field, the
temperature $T$ happens to be $T_{ep}$ i.e. $T_{ep} \sim
\frac{m_{\phi}}{\sqrt{\lambda}}$. Upon that,
\begin{equation}
N\simeq \ln(\frac{\lambda^{1/2}T_{initial}}{m_{\phi}}) = \frac{1}{2}\ln(\frac{\lambda T_{initial}^{2}}{m_{\phi}^{2}})
\end{equation}
This equation is exactly the same form as in new old inflation\cite{Dvali:2003vv}, because we pick $T_{ep}$ a special value here.
\vskip 9mm

\section{Conclusions}
Designing successful models of inflation within the frame of
supergravity and string theory seems to be a difficult task, since the
various moduli have protected masses $m$ of order $H$, the Hubble
constant during inflation, which breaks the slow roll condition
required for inflation. The prospect of avoiding the fine tuning
associated with slow roll inflation is very appealing. Dvali and
Kachru have proposed the idea of locked
inflation\cite{Dvali:2003vv,Dvali:2003us}. The motivation behind
the proposal is perhaps reinforced when we take on board the
recent suggestions concerning the string landscape, which invokes
the possibility that there are a discrete set of closely spaced
metastable vacua in string theory. However, there exist some
problems associated with saddle inflation, which renders the
original version of locked inflation not
feasible\cite{Easther:2004qs,Copeland:2005ey}.


We revisit the idea of locked inflation by asking whether there is
a possibility that we could make use of the merit while avoiding
the problems associated with saddle inflation. We propose a
scenario with an alternative mechanism of the evolution of the
``waterfall field''. Instead of letting it roll down the hill, we
use the tunneling of the $\phi$ field to end the locked inflation
stage, like in old inflation, by which the saddle inflation could
be avoided. Our model is based on the consideration of making locked inflation
feasible. Nevertheless it also can be seen as an effort to
embed the old inflation in string landscape. Further, we study the case of a cascade
with several stages of old locked inflation, which can be
motivated by the stringy landscape. Its relevance to other
inflation scenarios are also discussed.

\vskip 9mm

\section*{Acknowledgements}
The authors thank Yi-Fu Cai and Ya-Juan Zheng for helpful discussions and comments.
This work is supported in part by NSFC under Grant No: 10775180,
in part by CAS under Grant No: KJCX3-SYW-N2.
\begin{appendix}
\section{A Toy Model \label{appendix:A}}

To give
an explicit illustration, we need a specific formulae of the barrier.
While, string theory is not yet fully formulated, and we cannot yet give a detailed
discription of the potential but only some general features.
Anyway, let's consider a particular case that the barrier along $\phi$
direction is a symmetric function (the dashed line in Fig.1),
\begin{equation}
U(\phi)=\frac{\alpha}{16}((\phi^{2}-\frac{\mu}{\alpha^{\frac{1}{2}}})^{2}-\frac{\mu^{2}}{\alpha})^{2}=\frac{\alpha}{16}(\phi^{2}-2\frac{\mu}{\alpha^{\frac{1}{2}}}\phi)^{2},
\end{equation}
which has two minima $U^{'}(0)=U^{'}(\frac{\mu}{\alpha^{\frac{1}{2}}})=0$, and $U^{''}(0)=U^{''}(\frac{\mu}{\alpha^{\frac{1}{2}}})=\frac{1}{2}\mu^{2}$
where
\begin{equation}
\frac{1}{2}\mu^{2} \equiv \frac{1}{2}m_{\phi}^{2}+ \frac{1}{2}\lambda \Phi^{2}
\end{equation}
$\mu^{2}$ is defined as the effective mass squared of $\phi$ field.
Note that, just as we expected, both the height ($\frac{\mu^{4}}{16\alpha}$) and the width ($\frac{2\mu}{\alpha\frac{1}{2}}$) of the barrier will increase as $\Phi$ augments.
Then we add a term that breaks the symmetry,
\begin{equation}
V=U-\frac{\alpha^{\frac{1}{2}}\epsilon }{2\mu}\phi=\frac{1}{16}(\alpha\phi^{4}-4\alpha^{\frac{1}{2}}\mu\phi^{3}+4\mu^{2}\phi^{2})-\frac{\epsilon\alpha^{\frac{1}{2}} }{2\mu}\phi
\end{equation}
This potential is illustrated in Fig.1 (the thick line) as well as Fig.2.

\begin{figure}[htp]\centering
\begin{tabular}{cc}
  \includegraphics{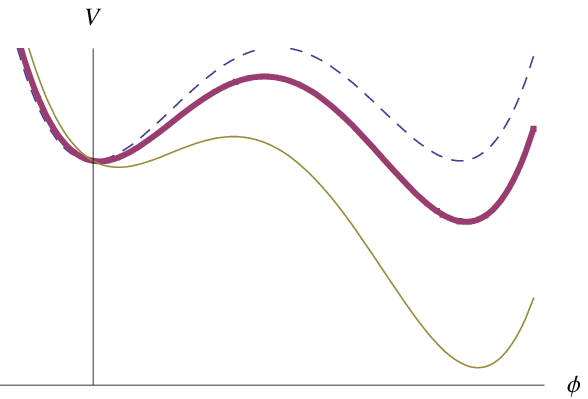}  & \includegraphics[scale=0.6]{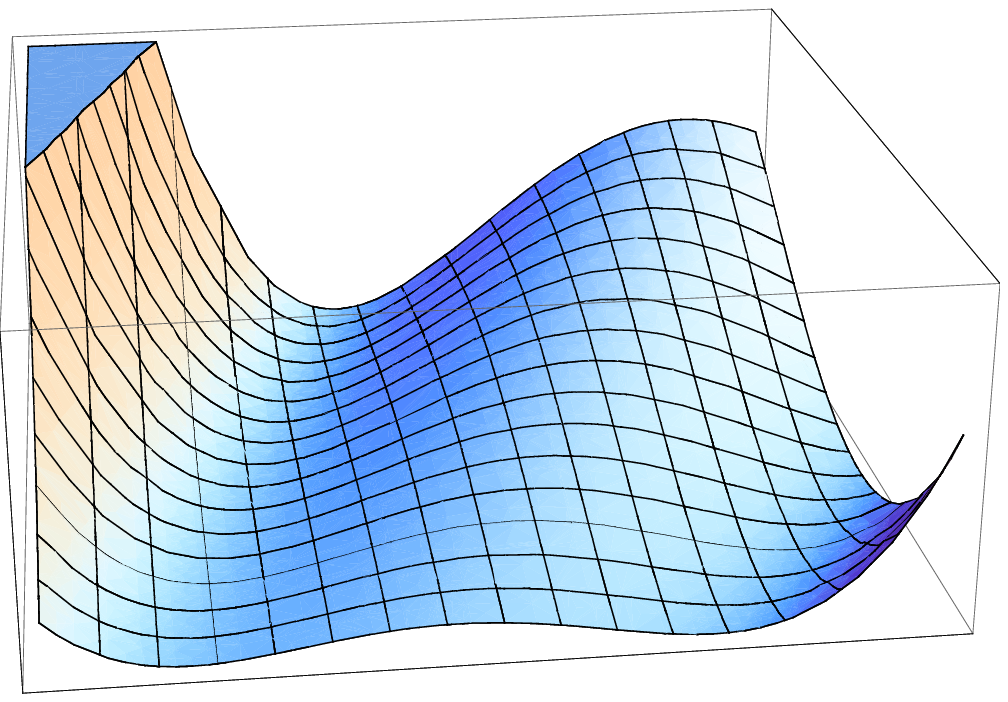}
\end{tabular}

\caption{A symmetric potential (dashed line) with a small symmetry breaking term leads to the potential of the purple form. The energy difference between the two minima is $\epsilon$. The lowest line represents the case of a large $\epsilon$. The direction of $\Phi$ field is perpendicular to this diagram.}
\label{Fig1}
\caption{A panoramic view of the potential in $\phi-\Phi$ plane (Fig.1 is a slice of it). It demonstrates that both the width and the height of the barrier will increase as $\Phi$ augments.}
\label{Fig2}
\end{figure}

And the approximate expression for B is
\begin{equation}
B=\frac{27\pi^{2}S_{1}^{4}}{2\epsilon^{3}}\simeq\frac{\pi^{2}\mu^{12}}{6\epsilon^{3}\alpha^{4}},
\end{equation}
and we draw the diagram of $\Gamma/H^{4}$ as the function of $\Phi$ in Fig.3\footnote{To draw this diagram we take $\epsilon = m_{\phi}^{4}$ as a small constant value (It's $\sim 10^{-31}\Lambda_{0}$, $\Lambda_{o}$ denotes the vacuum energy density of $H\sim TeV$). Though it is small, $\epsilon$ will exceeds the height of the barrier ($\frac{\mu^{4}}{16\alpha}$) when $\mu$ is extremely small. This is like the case of adding a relatively large symmetry breaking term (as the lowest curve in Fig.1 shows) and would break the thin wall approximation, which means the equation we used in this region is incomplete, but it won't change our qualitative result since $\Gamma$ is definitely in the inverse ratio of the height and width of the barrier and is directly proportional to $\epsilon$.}.

\begin{figure}[htp]\centering
  \includegraphics[height=8cm,width=10cm]{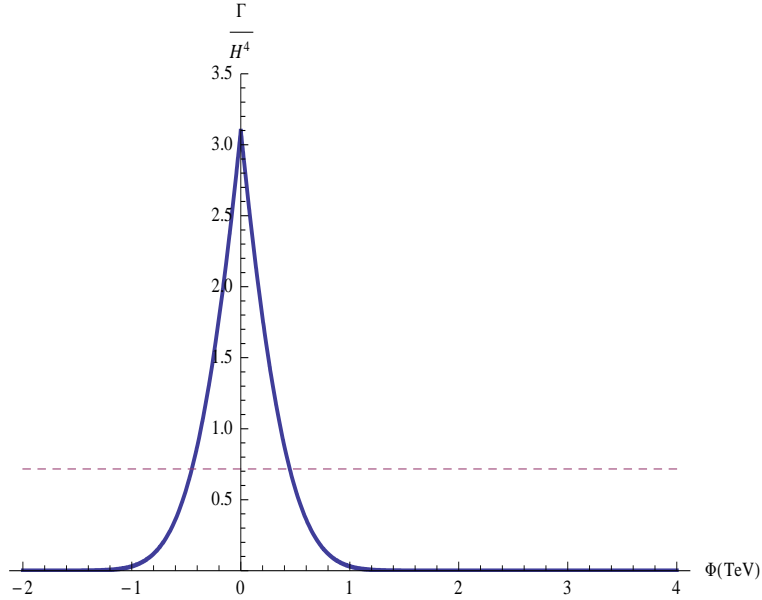}
\caption{$\Gamma/H^{4}$ as the function of $\Phi$. The dashed line stands for the critical value $\Gamma/H^{4}=9/4\pi$.
We take values of $H\simeq TeV$, $m_{\phi}\simeq 2H$ and $\alpha=\lambda=1$. It shows that as $\Phi$ decreases $\Gamma/H^{4}$ remains extremely small
and then increases suddenly, which guarantee the universe nucleated and thermalized throughout, and the large void problem will also be settled.}
\label{Fig3}
\end{figure}

From the diagram, we see that in this special example $\Phi_{ep}\simeq 0.5TeV\simeq\frac{1}{4}m_{\phi}$. And for certain potential with $\Phi_{ep}\simeq \frac{1}{m}\frac{m_{\phi}}{\sqrt{\lambda}}$ we will get $\frac{2}{3}\ln m$ additional e-foldings.
If we consider other form of potential, the barrier along $\phi$ direction is the form of
trigonometric function, and the height of it is
some kind of proportional to the value of $\Phi$ field ($U(\Phi,\phi)=U_{0}(\Phi)\sin\phi$) for example, we could also get the similar results.

\vskip 3mm
\section{Time Dependence Of The Expansion Rate \label{appendix:B}}

The reason why we take the initial value of $\Phi$ field to be
$\Phi_{\textit{initial}}\sim M_{p}$ is not only $M_{p}$ is a
natural characteristic value but also the following ones:
As the oscillation takes place in the plane of $\phi=0$
with an energy of $\frac{1}{2}m_{\Phi}^{2}\Phi^{2}$,
the Friedmann Equation becomes
\begin{equation}
\frac{3}{8\pi}H^{2}M_{p}^{2}= V(0,0)+\frac{1}{2}m_{\Phi}^{2}\Phi^{2} \simeq V(0,0)+\frac{1}{2}H^{2}\Phi^{2},
\end{equation}
the approximation is because of $m_{\Phi}\sim H$.
According to this equation we provide Tab.1.

\begin{table*}[htbp]
 \begin{center}

\begin{tabular}{|c|c|}
\hline
$\Phi (M_{p})$    &   ${V(0,0)}/\frac{1}{2}m_{\Phi}^{2}\Phi^{2}$\\
\hline
 0.4 &  1 \\
\hline
 0.1 &  25\\
\hline
$10^{-2}$ &  2500\\
\hline
$10^{-10}$& $2.5 \times 10^{19}$\\
\hline

\end{tabular}
\caption{The ratio of potential to kinetic energy in the case of different amplitudes of $\Phi$.
When $\Phi$=$\alpha M_{p}$, $V(0,0)/\frac{1}{2}m_{\Phi}^{2}\Phi^{2}=\frac{1}{4\alpha^{2}}-1$.
The potential energy becomes dominant as the amplitude of $\Phi$ decreases.}

\end{center}
\end{table*}

The table indicates that as long as $\Phi_{\textit{initial}}\lesssim M_{p}$,
the Hubble parameter will basically remain almost constant throughout
although the field $\Phi$ changes its amplitude by
many $\sim 17$ orders during the inflation.

Anyway, in the event that $\Phi_{\textit{initial}} \gtrsim M_{p}$,
with the stage of $\frac{1}{2} m_{\Phi}^{2}\Phi^{2} > V(0,0)$ at the beginning the
$\Phi$-oscillation dominates and there would be an interval of matter-dominated
expansion until the amplitude of $\Phi$ decreases to $\sqrt{2V(0,0)}/m_{\Phi}$,
the energy of $\Phi$ becomes subdominant to $V(0,0)$.
In this case, the hubble parameter first decreases approximately $\sim \frac{2}{3t}$,
then remains approximately constant.

\end{appendix}

\end{document}